# In-Depth Behavior Understanding and Use: The Behavior Informatics Approach

Longbing Cao

*Faculty of Engineering and Information Technology,*
*University of Technology Sydney, Australia*

**Abstract**

The in-depth analysis of human behavior has been increasingly recognized as a crucial means for disclosing interior driving forces, causes and impact on businesses in handling many challenging issues such as behavior modeling and analysis in virtual organizations, web community analysis, counter-terrorism and stopping crime. The modeling and analysis of behaviors in virtual organizations is an open area. Traditional behavior modeling mainly relies on qualitative methods from behavioral science and social science perspectives. On the other hand, so-called behavior analysis is actually based on human demographic and business usage data, such as churn prediction in the telecommunication industry, in which behavior-oriented elements are hidden in routinely collected transactional data. As a result, it is ineffective or even impossible to deeply scrutinize native behavior intention, lifecycle and impact on complex problems and business issues. In this paper, we propose the approach of *Behavior Informatics* (BI), in order to support explicit and quantitative behavior involvement through a conversion from source data to behavioral data, and further conduct genuine analysis of behavior patterns and impacts. BI consists of key components including *behavior representation*, *behavioral data construction*, *behavior impact analysis*, *behavior pattern analysis*, *behavior simulation*, and *behavior presentation* and *behavior use*. We discuss the concepts of behavior and an abstract behavioral model, as well as the research tasks, process and theoretical underpinnings of BI. Two real-world case studies are demonstrated to illustrate the use of BI in dealing with complex enterprise problems, namely analyzing exceptional market microstructure behavior for market surveillance and mining for high-impact behavior patterns in social security data for governmental debt prevention. Substantial experiments have shown that BI has the potential to greatly complement the existing empirical and specific means by finding deeper and more informative patterns leading to greater in-depth behavior understanding. BI creates new directions and means to enhance the quantitative, formal and systematic modeling and analysis of behaviors in both physical and virtual organizations.

## 1. Introduction

Behavior is the action or reaction of an entity, human or otherwise, to situations or stimuli in its environment. It can be widely seen anywhere at any time. In Google, the keyword 'behavior' returns 28,100,000 items while 'behavior analysis' hits 20,200,000 results. It is a key entity in understanding the driving forces and cause-effects of many issues. Besides the widespread and long-standing studies from a social and behavioral science perspective, behavior has also been increasingly highlighted for complex problem-solving within virtual and physical organizations, in particular for pattern analysis and business intelligence in many areas such as multi-agent organizations [16, 7], customer relationship management [36], fraud detection [24, 38], outlier detection [1, 28, 34], intrusion detection [43], social computing [44], web usage [23, 41], and network monitoring [25]. In customer relationship management, for instance, it is widely agreed that customer behavior analysis is essential for deeply understanding and caring for customers, and eventually for boosting enterprise operation and enhancing business intelligence. Other typical examples include churn analysis of telecommunication customers from one provider to another [2, 36, 37], credit estimation of banking customers for home loans and finance transactions [3, 48, 18], exceptional behavior analysis of terrorists and criminals [17], and trading pattern analysis of investors in capital markets [47, 21, 12].

To the best of our knowledge, behavior analysis has usually been directly conducted on customer demographic and transactional data. For instance, in telecom churn analysis, service subscriber demographics and their usage, billing, credit, application and complaint history are analyzed to classify customers into loyal (stayer) and non-loyal (churner) groups based on the dynamics of usage change [2, 36, 37]; while in outlier mining of trading behavior, price movement is usually focused on detecting abnormal trading [21, 47]. In scrutinizing the datasets used in the above examples, we realize that the so-called behavior analysis is actually not based on genuine behavioral elements, rather on straightforward customer demographic data and business usage related transactions (for simplicity, called 'transactional data') accumulated during business operation.

In general, customer demographic and transactional data is not organized in terms of behavior but on entity relationships. Entities and their relationships collected in transactions reflect those objects closely related to particular business problems. For instance, in stock markets, orderbook transactions mainly record and manage price, volume, value and index related information. In transactional data, behavior is hidden and behavioral properties are split and separately recorded. Consequently, behaviors are *implicitly* and *dispersedly* recorded in transactional data. Such *behavior implication* indicates the limitation or even ineffectiveness of supporting genuine behavior analysis, principally for the following reasons.

- First, the behavior implication in source data including transactions determines that it cannot support in-depth analysis on *behavior interior* supported by behavioral elements, rather on the result of *behavior exterior*, where business appearance (such as stock market price movement) is focused, actually hiding and excluding behavioral elements.
- Second, with behavior implied in transactional data, it is not possible to scrutinize behavioral intention and impact on business appearance and problems; whilst behavior may play an important role in the appearance of problems, this role has been weakened or even ignored as a potential factor in traditional customer behavior analysis.

A question to be asked is why and how behavioral data makes a difference in pattern analysis and business intelligence.

- First, in many cases, behavior plays the role of an internal driving force or cause for business appearance and problems. Most business problems, such as mobile customer churning, can be better understood and investigated if customer behavior can be focused and scrutinized.

- Second, when behavior is disclosed and taken as an extra factor in problem-solving, it can greatly complement traditional pattern analysis that relies solely on demographic and transactional data, and can disclose extra information and relationships between behavior and target business problems. In this way, a multi-dimensional viewpoint and solution may exist that can uncover problem-solving evidence from not only demographic and transactional perspectives, but also from a behavioral perspective (including intentional, social and impact aspects). As a result, the identified patterns combine multiple aspects of information and thus are more informative for problem understanding and solving, which cannot be achieved directly through traditional approaches.

In order to support genuine analysis on behavior interior, it is essential to make behavior recorded in data storage 'explicit' by squeezing out behavioral elements hidden in source data including transactions and relevant business management information systems. For this, *a data conversion from transactional space to behavior feature space is necessary*. The conversion extracts, transforms and presents behavioral elements, and reorganizes them into behavioral data that caters for behavior analysis. This is the process of *behavior modeling and mapping*. As a result, in behavioral data, behavior is *explicit*, and is mainly organized in terms of behavior entities, properties and relationships. This leads to *behavior explication*. On the behavioral data, we can then explicitly and more effectively analyze behavior patterns and behavior impacts than on normal transactional data.

*Behavior modeling and representation*, *behavioral data construction*, *behavior impact modeling*, *behavior pattern analysis*, *behavior simulation*, *behavior presentation* and *behavior use* constitute the main goals and tasks of Behavior Informatics (BI) (or in this context, Behavior Computing). Building on the studies on mining activity data and activity sequential patterns [10, 11, 15], this paper presents an overall framework and key concepts of BI. Note that, given the limitation of the objectives, BI is proposed mainly from the perspectives of information technology and data analysis rather than from social and sociological aspects, in which human behavior has been intensively studied in different ways.

Let us continue to use the example of churn analysis of mobile customers to distinguish behavior analysis on behavioral data from traditional customer behavior analysis on transactional data. With BI, besides the analysis on demographic and business usage data, we can further analyze behavior sequences of a customer, from activities resulting from his/her registration and activation of a new account into a network, the distribution (such as frequency and duration) of making calls during an observation period, to the characteristics of making payments and the date of leaving the network. Obviously, analysis on such behavioral data can explore much more fruitful information about a mobile holder's intention, activity changes, usage dynamics, and payment profile than analysis simply on demographic and service usage data. This is essential for disclosing reasons for, and drivers of, customer churning and loyalty change.

We further illustrate the proposed concepts through two real-world case studies, namely analyzing market microstructure behavior patterns for market surveillance and social security activity patterns for overpayment prevention. Substantial experiments on real-world data from stock markets and an Australian government agency have shown the promising potential of BI for handling many critical application problems by combining behavior-oriented perspectives.

The paper is organized as follows. Related work is briefed in Section 2. In Section 3, the concepts of behavior and an abstract behavioral model are discussed. Section 4 introduces the framework of BI studies. Section 5 illustrates the analysis of exceptional market microstructure behavior patterns for market surveillance, while Section 6 demonstrates the discovery of high-impact behavior patterns in social security data for governmental debt prevention. We conclude the paper in Section 7.

## 2. Related Work

Behaviors can be seen everywhere in business and social life. Behavior-oriented modeling and analysis has been proposed and/or studied in areas such as human-machine interaction, data mining and machine learning. Typical concepts such as user modeling, activity monitoring, customer behavior analysis and web user behavior patterns can be found in a number of references.

User modeling is predominantly studied in human-computer interaction [32], which focuses on developing cognitive models of human users, including modeling of their skills and declarative knowledge for human-computer interaction and user testing. Even though machine learning can be used for modeling user behavior, its main objective is to observe user behavior in order to predict future user actions [45]. From here we can distinguish user modeling from behavior modeling by such aspects as objectives, means and outcomes. We will explain behavior modeling in Sections 3 and 4.

Activity monitoring has been proposed to monitor activities for unusual behavior [25, 46]. In web usage and preference analysis, web access information, weblogs and session information for navigational history, experience and location are mainly used to 'simulate' web user behavior [23, 26, 41]. Similarly, in online business customer behavior analysis, product awareness and exploration of purchase commitment are focused for pattern discovery [33]. Behavior profiles or models of user email accounts are analyzed in behavior-based email analysis [42].

Customer and consumer behavior analysis has been intensively inspected from marketing strategy and customer relationship management perspectives. For instance, researchers recognize the role of social factors in churn analysis in the mobile industry [19]. In capital markets, relational data related to brokers and securities are used to analyze security fraud [24, 38]. In analyzing market insider trading, price information is mainly focused on [21].

Additional techniques include context representation and reasoning [20, 30, 31], ontological engineering and semantic web [8]; sequence analysis [10] and reality mining [50, 22] are also helpful for behavior analysis. Contextual representation considers contextual information, while reality mining defines the collection of machine-sensed environmental data pertaining to human social behavior. With the increasing need to scrutinize behavior, other relevant techniques may also be helpful.

In current behavior analysis, static information is mainly focused, and behavior properties are often split and separately studied. There is not an explicit and systematic behavior view of dynamic, sequential, social and impact-oriented aspects. In addition, the existing approaches are mainly based on analyzing business appearances and unusual event occurrences.

Recently, *activity mining* [15, 11] has been proposed to analyze activity data, and its modeling, pattern analysis and impact analysis. Activities, events, actions, operations and interactions consist of the main embodiment of behavior. Preliminary research has been conducted on high impact activity sequence mining [10], exceptional behavior analysis [9], and microstructure-based market trading behavior analysis [12] in real-life data, including capital market data and social security data. This research uncovers a wide need and a range of applications for which systematically representing and analyzing behaviors would be desirable, namely establishing the informatics of behaviors.

## 3. An Empirical Behavioral Model

### *3.1. What Is Behavior About?*

Within the scope of Behavior Informatics (BI), *behaviors* refer to those activities that present as actions, operations or events as well as activity sequences conducted by entities within certain contexts and environments in either a virtual or physical organization. Even though behavior generally may also include other actions conducted by organisms such as animals, or more physical activities such as the movement of a robot, as discussed in social and behavioral sciences, we are particularly interested in behaviors recorded or converted into computational systems. From the source and computational perspectives, we categorize behaviors into *symbolic behavior* and *mapped behavior* from individual and group perspectives. In addition, we observe *individual behaviors* and *group behaviors*.

- 'Symbolic behavior' refers to social activities recorded into computational systems, which present as symbols, representing human interaction and operation with a particular object or object system. A typical example is a stock trader's behavior recorded in trading systems. For instance, an investor places an order into a trading system: here 'behavior' is to 'place an order'. Other examples include web user behavior, game user behavior and intelligent agent behavior. Typically, such human behavior happens in a certain social context, and therefore presents social characteristics.

Symbolic behavior is our main focus in BI. However, there is another type of behavior, namely mapped behavior, widely seen in areas such as computer vision and pattern recogni- tion. In this case, we see physical behaviors either directly or indirectly, which are often referred to as 'visual behavior'.

- 'Mapped behavior' refers to the images of physical activities recorded by sensors into computer systems, which present as the virtual version of an object's actions in the physical world. An example is human activity captured by video surveillance systems. Other examples include an agent's behavior, a robot's behavior and an organism's behavior in game systems. Such captured behavior is the direct or indirect mapping of physical behaviors into a virtual world.

With regard to mapped behaviors, our focus is on their pattern analysis rather than their detection, which is the main task of the relevant areas such as pattern recognition where the underlying behaviors are identified. Once behaviors are detected, they are further extracted and represented in terms of behavior representation methods into behavioral data, in a way similar to symbolic behaviors.

### *3.2. An Empirical Behavioral Model*

As an abstract concept, behavior ($\gamma$) presents many attributes and properties, for instance, action and action time. A behavioral model may be built to capture such attributes and properties. Based on empirical understanding of behavior in domains such as stock markets, we extract the following attributes to represent the general properties of behavior.

- Subject ($s$): The entity (or entities) that issues the activity or activity sequence;
- Object ($o$): The entity (or entities) on which a behavior is imposed;

- Context ($e$): The environment in which a behavior operates; context may include pre-condition and post-condition of a behavior;
- Goal ($g$): Goal represents the objectives that the behavior subject would like to accomplish or bring about;
- Belief ($b$): Belief represents the informational state and knowledge of the behavior subject about the world;
- Action ($a$): Action represents what the behavior subject has chosen to do or operate;
- Plan ($l$): Plans are sequences of actions that a behavior subject can perform to achieve one or more of its intentions;
- Impact ($f$): The results led by the execution of a behavior on its object or context;
- Constraint ($c$): Constraint represents what conditions impact on the behavior; constraints are instantiated into specific factors in a domain;
- Time ($t$): When a behavior occurs;
- Place ($w$): Where a behavior happens;
- Status ($u$): The stage where a behavior is currently located; for instance, status may refer to *passive* (not triggered), *active* (triggered but not finished yet) or *done* (finished); in some other cases, status may include *valid* or *invalid*;
- Associate ($m$): Other behavior instances or sequences of actions that are associated with the target; behavior associates possibly exist when a behavior has impact on another, or behaviors are related through interaction and business process to form a behavior network.

With the above behavioral attributes, a behavior can be represented in terms of a behavioral vector ($\vec{\gamma}$) as follows.

$$\vec{\gamma} = \{s, o, e, g, b, a, l, f, c, t, w, u, m\} \tag{1}$$

A behavior vector ($\vec{\gamma}$) not only consists of basic properties of a behavior such as *time* and *place*, but also social and organizational factors including *context*, *constraints*, and *impact*. It presents heterogeneous properties, which may consist of textual, categorical and numerical data.

Further, a behavior sequence ($\Gamma$) of a customer can be represented in terms of a vector sequence ($\vec{\Gamma}$), which consists of all behavior instances represented in vectors.

$$\vec{\Gamma} = \{\vec{\gamma_1}, \vec{\gamma_2}, \ldots, \vec{\gamma_n}\} \tag{2}$$

The above abstract behavioral model aims to capture the major features as a generic behavior object. It is worth noting that in deploying this abstract model into describing behaviors in different domains, some of those attributes and properties may not take place, while for other situations some of them may be embodied into specific features or multiple variables. Without loss of generality, a simplified behavioral model ($\vec{\gamma}$') is defined as follows:

$$\vec{\gamma}' = \{s, o, a, f, t\} \tag{3}$$

It indicates that a behavioral subject ($s$) conducts an action ($a$) on an object ($o$) at a time ($t$) which leads to a certain impact ($f$). Accordingly, a behavior sequence ($\vec{\Gamma}$') is as follows:

$$\vec{\Gamma}' = \{\vec{\gamma_1}', \vec{\gamma_2}', \ldots, \vec{\gamma_n}'\} \quad (4)$$

In Sections 5 and 6, we instantiate the behavioral model to represent real-life behaviors. The examples also show the customization and expansion of the abstract model to support domain-specific properties and/or attributes.

With the vector-based behavior sequences, further analysis of such vectors can identify vector-oriented patterns. Compared to traditional sequential pattern mining, such vector-oriented behavior pattern analysis is much more comprehensive and informative. To mine for patterns in such a complex data structure, it is not possible for existing data mining techniques to be directly deployed. One of the BI tasks is to study such vector-based behavior pattern analysis.

## 4. Framework of Behavior Informatics

### *4.1. Basic Concepts*

Behavior Informatics is a scientific field which aims to develop methodologies, techniques and practical tools for representing, modeling, analyzing, understanding and/or utilizing symbolic and/or mapped behavior, behavioral interaction and network, behavior patterns, behavior impacts, the formation of behavior-oriented groups and collective intelligence, and behavioral intelligence emergence. In essence, Behavior Informatics seeks to deliver computational technologies and tools for deeply understanding behavior and social behavior networks. In this sense, we also call it *behavioral computing*.

As a new research issue, BI consists of many open issues that are worthy of systematic investigation as well as case studies from aspects such as *behavioral data construction*, *behavior modeling and representation*, *behavior impact modeling*, *behavior pattern analysis*, and *behavior network analysis*. Additionally, *behavior measurement and evaluation*, *behavior presentation*, and *behavior use* are very important topics. In understanding behavior, *behavior simulation* can play an important role in both artificial systems and real societies. We further expand these by listing some key research issues for each of the above research topics, but certainly there may be other issues.

(1) *Behavioral Data Construction*: In many cases, it may be necessary to convert normal source data into behavior-oriented feature space, in which behavior elements constitute the major proportion of the dataset. Research issues may consist of behavioral feature selection, mapping from source data to behavioral data, behavioral data transformation, or quality issues in behavioral data.

(2) *Behavior Modeling and Representation*: Refers to the modeling and representation of behaviors by developing representation languages and tools. This is to build formal methods and techniques to capture behavioral entities and their attributes and properties, as well as to represent relationships between behavior entities. The modeling techniques can also be used to understand interaction, causality, convergence, divergence, selection, decision, evolution and emergence of behavior entities, behavior networks and behavior impact. For these, modeling languages, specifications and tools need to be developed.

(3) *Behavior Impact Analysis*: Behavior having an impact on business, politics and societies is our major interest. This means that the impact of behavior and the behavior network may be on economic, cultural, organizational, social and political aspects. To analyze behavior impact, techniques such as impact modeling, measurements for risk, cost and trust analysis, the transfer of behavior impact under different situations, and exceptional behavior impact analysis would be very helpful. The analytical results will be utilized for detec- tion, prediction, intervention and prevention of negative behavior or for opportunity use if positive cases are identified. Research issues may include behavior impact modeling,organizational/social impact analysis, risk, cost and trust analysis of behavior, impact transfer process and patterns, opportunity analysis, detection and prediction.

(4) Behavior network: Multiple sources of behavior may form into certain behavior networks. Particular human behavior is normally embedded into such networks to fulfill its roles and effects in a particular situation. Behavior network analysis aims to understand the intrinsic mechanisms inside a network; for instance, behavioral rules, interaction protocols, convergence and divergence of associated behavioral itemsets, as well as their effects such as network topological structures, linkage relationships, and impact dynamics.

(5) *Behavior Pattern Analysis*: This is the major focus of BI: to identify patterns in behavior entities and behavior networks. For this, we need first to understand behavior structures, semantics and dynamics in order to further explore behavior patterns. We then investigate pattern analytical tasks such as detection, prediction and prevention of critical behavior, misbehavior, and behavior impact through utilizing and inventing approaches such as activity mining, correlation analysis, linkage analysis, clustering and combined pattern mining. Issues may consist of hidden and emergent behavioral structures, behavior convergence and divergence patterns, behavior stream mining, dynamic behavior pattern analysis, visual behavior pattern analysis, cause-effect analysis, parallel/sequential behavior sequences, multiple sequence analysis, demographic-behavioral combined pattern analysis, social networking behavior, misbehavior detection and prediction in communities, behavior self-organization, evolution and emergence, contextual behavior network analysis and mining, and exceptions and outlier mining.

(6) *Behavior Simulation*: Simulation can play an essential role in deep understanding of behavior working mechanisms, interaction amongst behavior instances, dynamics and the formation of behavior group and behavior intelligence emergence. For example, simulation can be conducted on a large-scale behavior network, convergence and divergence, evolution and adaptation of behavior, group behavior formation and evolution, hidden/explicit community analysis and formation, artificial behavior impact analysis system through setting up artificial and computation-oriented behavior systems. In addition, it is also interesting and important to model the interaction between behavior subjects and their surrounding environment, and to analyze the impact of key organizational and social factors. A potential tool and approach for behavior simulation is multiagent-based behavior network modeling and analysis.

(7) *Measurement and Evaluation*: To measure the impact of behavior and behavior networks, and to measure the significance of behavior patterns, appropriate measurement and evaluation systems need to be developed. This includes not only general efforts on developing technical measures, but also particular work on business performance and effect that can be measured in terms of risks, benefits and costs from economic, organizational, operational, social and political perspectives. Specific issues may also feature in evaluating behavior patterns and impacts such as utility, privacy, security, scalability, reliability and actionabil-ity.

(8) *Behavior Presentation*: The presentation of the dynamics of behavior and the behavior network in varying aspects would assist with the understanding of the behavior lifecycle and impact delivery. Equally, proper presentation of resulting behavior analysis is important for end users and domain experts to enable them to take over and act on business decisions. Research issues may include rule-based behavior presentation, visualization of the behavior network, visual analysis of behavior patterns, flow and network visualization, sequence visualization, dynamic factor tuning, configuration and effect analysis, and distributed, linkage and collaborative visualization.

(9) *Behavior Use*: Besides the aim of disclosing the dynamics, working mechanisms and potential impact of behavior and behavior networks, BI is also particularly interested in the use of resulting behavior patterns and analytical outcomes. In this regard, many things may be investigated, for instance, personalization and customization of customer services and relationship management, detection, prediction and intervention of critical events/actions/operations and networks, recommendations for product sale, service provision and information acquisition, and the establishment and formation of critical organizational/social relations, groups and communities.

Figure 1 further illustrates major research tasks/approaches and the relations among the above key research components. Behavioral data is extracted from behavior-relevant applications, and then converted into a behavioral feature space. When the behavioral data is ready, behavior pattern analysis and impact analysis are conducted on the data. To support behavior pattern analysis and impact analysis effectively, behavior simulation and modeling can provide fundamental results about behavior dynamics and relevant businesses and tools for knowledge discovery. Besides supplying another point of view for behavior analysis, behavior presentation contributes the techniques and means to describe and present behavior.

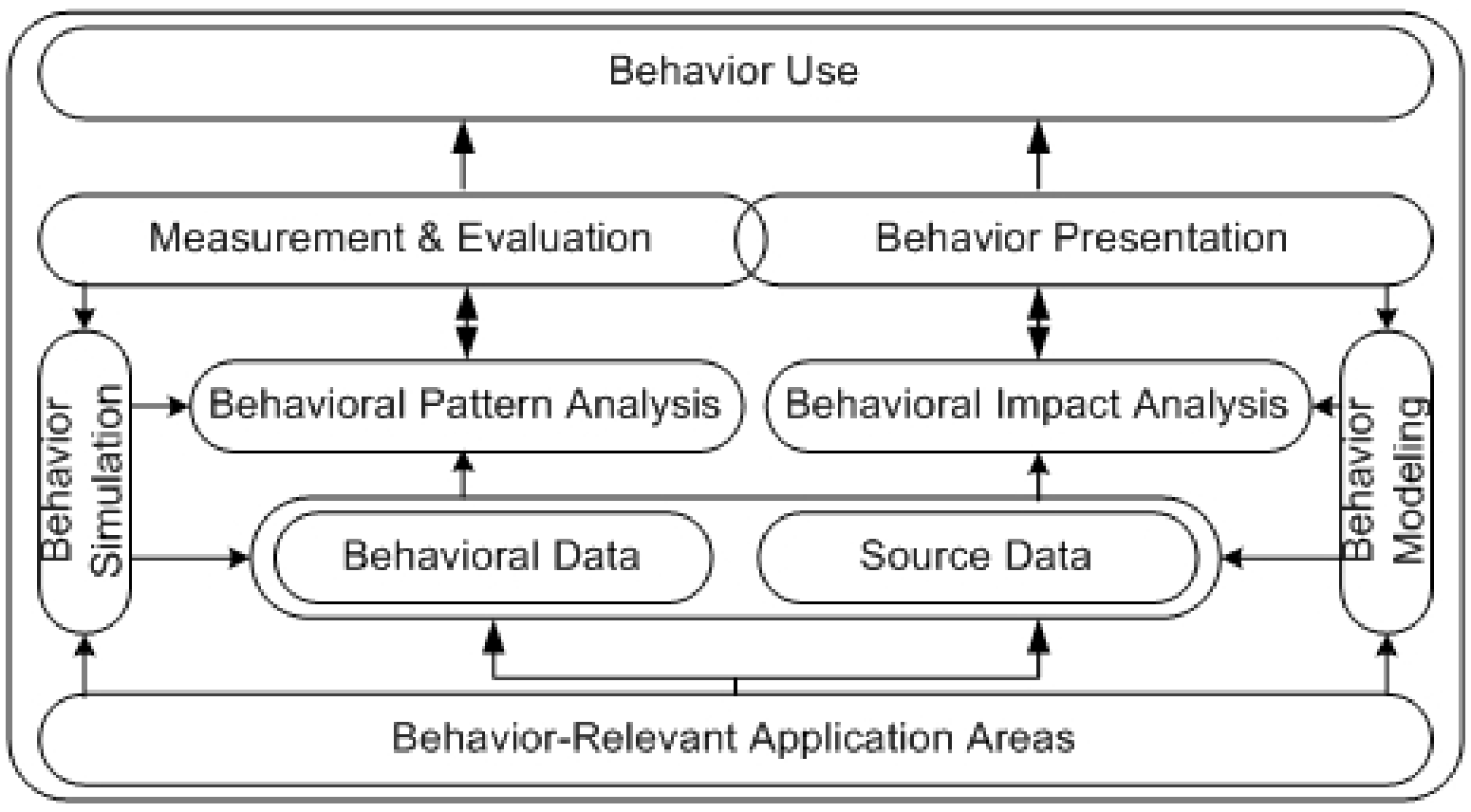

Figure 1: Research Map of Behavior Informatics

### 4.2. Basic Process

With the BI components discussed above, we can sketch a generic process of BI as shown in Figure 2. From the concept perspective, BI is a process converting entity relationships (*DB*) oriented transactional data ($\Psi$) to behavior feature-oriented data ($\vec{\Gamma}$) through behavior modeling ($\Theta(\vec{\Gamma})$), analyzing behavior patterns ($P(\vec{\Gamma})$) and impacts ($I(\vec{\Gamma})$) in terms of developing behavior pattern mining methods ($\Omega$), presenting behavior patterns ($V(\vec{\Gamma})$), and transforming into decision-support business rules ($\tilde{R}$). The outcomes of BI consist of behavior patterns ($\tilde{P}$) and further corresponding business rules ($\tilde{R}$) for business decision-making.

$$BI : \Psi(DB) \xrightarrow{\Theta(\Gamma)} \vec{\Gamma} \xrightarrow{\Omega,e,c,t_i()} \widetilde{P} \xrightarrow{\Lambda,e,c,b_i()} \widetilde{R} \tag{5}$$

Following the principle of actionable knowledge discovery [6], this process can be further decomposed and modeled in terms of the following steps.

---

BI PROCESS: The Process of Behavior Informatics

INPUT: original dataset $\Psi$;
OUTPUT: behavior patterns $\tilde{P}$ and business rules $\tilde{R}$;

Step 1: Behavior modeling $\Theta(\vec{\Gamma})$;
    Given dataset $\Psi$;
    Develop behavior modeling method $\theta$ ($\theta \in \Theta$) with technical interestingness $t_i()$;
    Employ method $\theta$ on the dataset $\Psi$;
    Construct behavior vector set $\vec{\Gamma}$;
Step 2: Converting to behavioral data $\Phi(\tilde{\Gamma})$;
    Given behavior modeling method $\theta$;
    FOR $j$ = 1 to ($count(\Psi)$)
        Deploy behavior modeling method $\theta$ on dataset $\Psi$;
        Construct behavior vector $\vec{\gamma}$;
    ENDFOR
    Construct behavioral dataset $\Phi(\vec{\Gamma})$;
Step 3: Analyzing behavior patterns $P(\vec{\Gamma})$;
    Given behavioral data $\Phi(\vec{\Gamma})$;
    Design pattern mining method $\omega$ ($\omega \in \mathbf{\Omega}$); Employ the method $\omega$ on dataset $\Phi(\vec{\Gamma})$;
    Extract behavior pattern set $\widetilde{P}$;
Step 4: Converting behavior patterns $\widetilde{P}$ to business rules $\widetilde{R}$
    Given behavior pattern set $\tilde{P}$;
    Develop behavior modeling method $\mathbf{\Lambda}$;
    Involve business interestingness $b_i()$ and constraints $c$ in the environment $e$;
    Generate business rules $\widetilde{R}$

---

*4.3. Theoretical Underpinnings*

Behavior Informatics is a multidisciplinary research field. Its theoretical underpinnings involve analytical, computational and social sciences as shown in Figure 3. We interpret the theoretical infrastructure for BI from the following perspectives: (1) Theoretical foundation, (2) Fundamental technologies, and (3) Supporting techniques and tools.

From the *theoretical foundation* perspective, BI draws theoretical support from multiple disciplines, including mathematics, information sciences, intelligence sciences, system sciences, cognitive sciences, psychology, social sciences and sciences of complexities. Mathematics provides formal methods for modeling behavior. Information and intelligence sciences provide support for intelligent information processing and systems. System sciences furnish methodologies and techniques for behavior and behavior network modeling and system simulation, and for large scale behavior networks. Cognitive sciences incorporate principles and methods for understanding human behavior belief, the intention and goal of human behavior. Psychology can play an important role in understanding human behavior motivation and evolution. Social sciences supply foundations for conceiving organizational and social factors and business processes surrounding behavior and embedded in behavior networks. Areas such as economics and finance are also important for understanding and measuring behavior impact. Methodologies from the science of complexities is essential for group behavior formation and evolution, behavior self-organization, convergence and divergence, and behavior intelligence emergence.

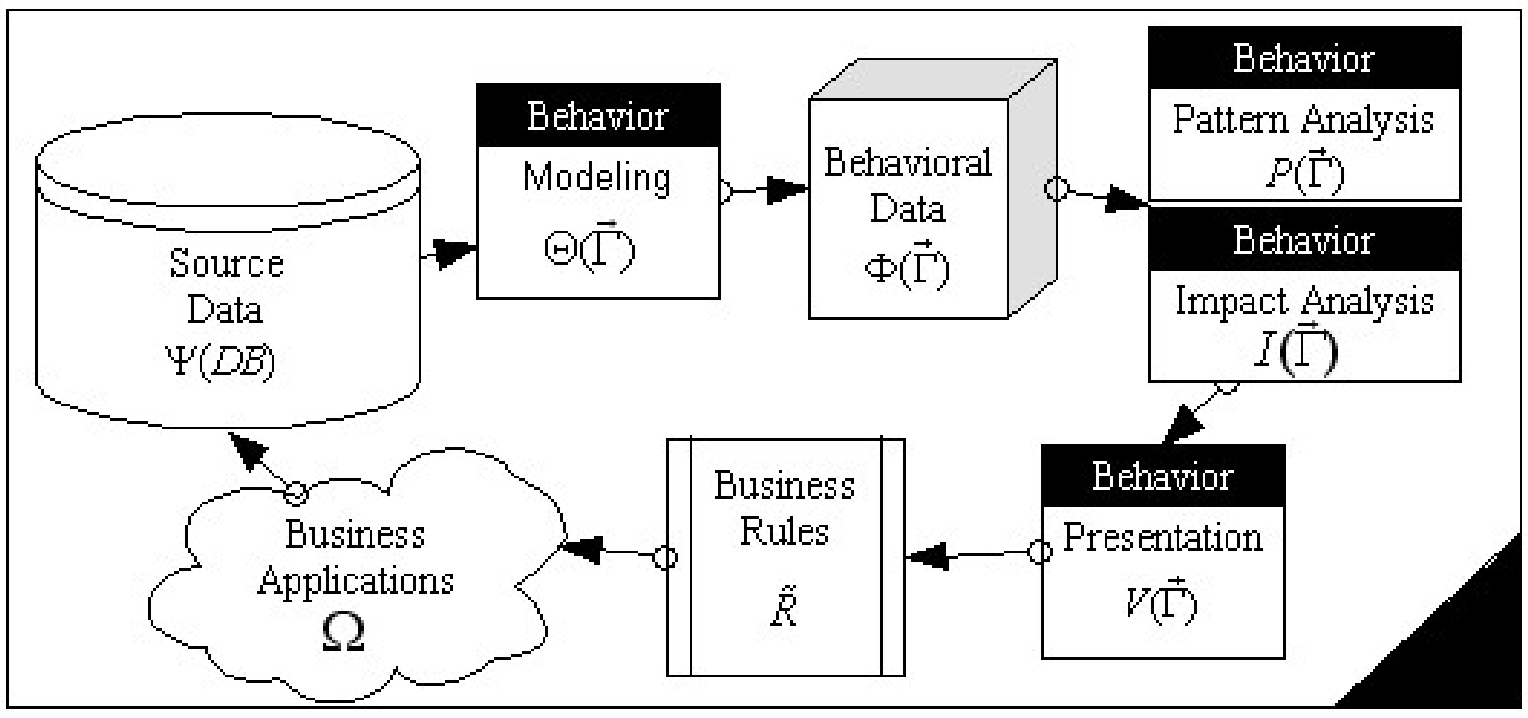

Figure 2: The Behavior Informatics Process

*Fundamental technologies* are necessary for behavioral modeling, pattern analysis, impact analysis, and behavior simulation. To support behavior modeling, technologies such as user modeling, formal methods, logics, finite state machine, context representation, ontological engineering, semantic web, group formation and cognitive science are essential. They can not only represent behavioral elements, but also contribute to the mapping from transactional entity space to behavioral feature space. The modeling of behavior impact needs to refer to technologies in areas such as risk management and analysis, organizational theory, sociology, psychology, economics and finance. For the analysis of behavior patterns, technologies such as data min- ing and knowledge discovery, artificial intelligence and machine learning can contribute a great deal. In simulating behavior, behavior impact and behavior networks, we refer to techniques and tools in fields like system simulation, artificial social system, open complex systems, multiagent systems, swarm intelligence, social network analysis, reasoning and learning. The presentation of behavior evolution and behavior patterns can benefit from areas of visualization and graph theory.

From the operationalization aspect, BI needs to develop effective techniques and tools for representing, modeling, analyzing, presenting and/or utilizing behavior. This involves many specific approaches and means. For instance, several methods such as algebra and logics may be useful for modeling behavior. The behavior pattern analysis may involve many existing tools such as classification and sequence analysis, as well as the development of new approaches. To simulate behavior impact, one may use agent-based methods for cause-effect analysis, while for presenting behavior, visualization techniques may be useful.

Currently, there are many emergent advances in information sciences and intelligence sci-ences. Some typical fields relevant to BI include *agent computing* [1], *service computing* [2], *organizational computing*, *social computing* [3], and *human-centered computing*. These new computing paradigms are all related and can greatly contribute to Behavior Computing as discussed in this paper. In the context of BI, some sort of interaction and integration [4] of relevant computing paradigms can certainly contribute to the problem-solving. A possible path for thecollaboration and integration of the above computing paradigms for BI is as follows: organizational computing, social computing and human-centered computing serve for behavior and behavior network, impact and pattern representation, analysis and evaluation, agent computing contributes to behavior simulation, and service computing contributes to behavior use. To support the integration and metasynthesis, the theory of *M-Space* and its working mechanisms *M-Interaction* and *M-Computing* [4] can provide solutions and directions.

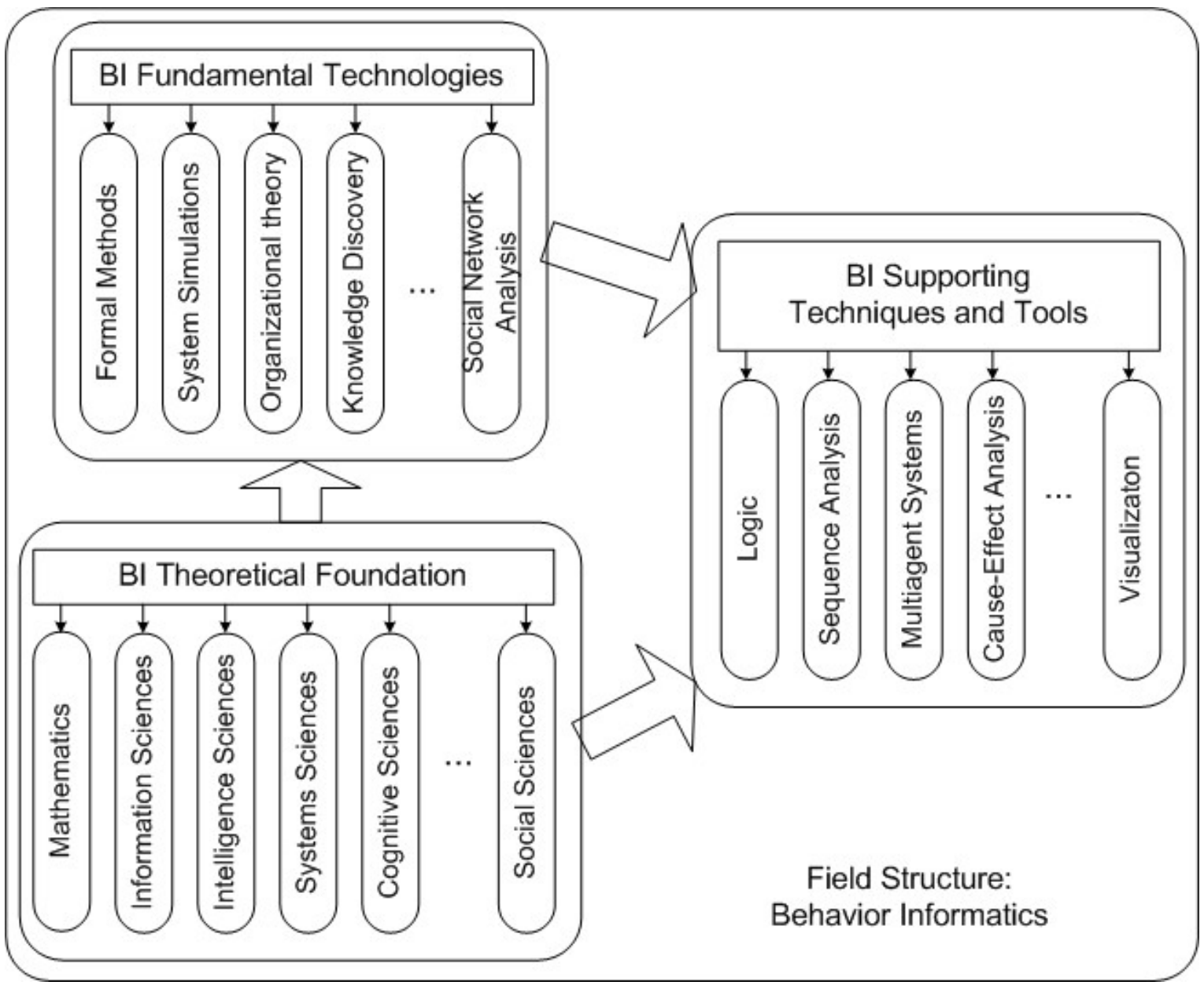

Figure 3: Field Structure of Behavior Informatics

## 5. Case Study 1: Market Microstructure Behavior Analysis

### *5.1. Market Microstructure Behavior in Capital Markets*

In capital markets, trading behavior refers to actions and operations conducted by investors and recorded into trading engine systems. Basic trading actions consist of buy (*B*), sell (*S* ) and hold (*H*). Operations conducted by investors include the interaction and intervention activities between investors and trading engines; for instance, placing an order or withdrawing an order. Such trading behavior provides significant information about the investment intention and corresponding actions taken by investors to achieve that intention, for example, manipulating orders to take advantage of the resulting manipulated market movement.

[1]International Conference on Autonomous Agents and MultiAgent Systems.

[2]International Conference on Service Oriented Computing.

[3]International Conference on Social Computing.

[4]Readers may refer to the Special Interest Group of Agent-Mining Interaction and Integration: www.agentmining.org or references [13, 14] for more information on computing paradigm integration.

Table 1: Trade order sequences related to the order O100

| Serial ID | Date | Time | Account ID | Security | Action | Price | Volume |
|---|---|---|---|---|---|---|---|
| O100 | 28/06/2005 | 09:54:07 | A123 | S123 | B | 10.00 | 1000 |
| O078 | 28/06/2005 | 09:59:52 | A348 | S123 | S | 10.10 | 500 |
| O102 | 28/06/2005 | 09:59:52 | A980 | S123 | B | 10.10 | 500 |
| O067 | 28/06/2005 | 09:59:56 | A690 | S123 | S | 10.00 | 200 |
| O100 | 28/06/2005 | 09:59:56 | A123 | S123 | B | 10.00 | 200 |
| O089 | 28/06/2005 | 10:07:49 | A531 | S123 | S | 10.00 | 300 |
| O100 | 28/06/2005 | 10:07:49 | A123 | S123 | B | 10.00 | 300 |

In capital markets, the above trading behavior is governed by the theory of *market microstructure* [27, 35]. Consequently, trading engine systems record relevant information such as orders and trades into transactions, which consist of *market microstructure data* [12]. In general, market microstructure data consists of basic data categories including orders, trades, indices and market data. It involves several dimensions such as time, value and trade actions.

The existing research in capital markets, from either finance, such as behavior finance [39], or information technology areas like insider trading analysis [21], mainly rely on data related to market and price movements. To the best of our knowledge, little research has been done on analyzing genuine trading behavior to disclose interior driving forces and causes of market movement and related investment performance.

In fact, transactions in market microstructure data reflect the results of investors' trading behavior in a market, which is controlled by certain trading models following market microstructure theory. We call such trading behavior that is monitored in terms of market microstructure theory *market microstructure behavior*. Though such transactions are usually managed in terms of entities such as orders and trades, we can actually scrutinize market microstructure behavior hidden in the transactions by squeezing out behavioral elements, and then make them explicit to form behavioral data for behavior-oriented analysis. This is to effect the *shift from behavior implication in transactional space to explication in behavioral data*. For this, behavior modeling is necessary for extracting and representing market microstructure behavior in market transactions.

### *5.2. Modeling Market Microstructure Behavior to Construct Microstructure Behavioral Data*

Normal orderbook transactions available from stock markets consist of attributes from related entities such as *orders* and *trades*. Table 1 illustrates a data sample of stock transactional data. Such data somehow indicates an investor's actions such as whether a trade is a buy (*B*) or a sell (*S* ), while it does not explicitly exhibit an investor's intention and corresponding behavior in the market. For such a purpose, we have to convert the source data including transactions into microstructure behavioral data.

Following the vector-based behavioral model discussed in section 3, we build vector-based microstructure behavior sequences to model market microstructure behavior. Considering the fact that every order follows market microstructure theory and indicates information about the order holder's intention, the proper representation of a trade behavior instance should reflect the trader's intention and actions associated with corresponding order and trade lifecycles.

In constructing the trading behavior sequences in terms of the abstract behavioral model $\vec{\gamma}$, we only consider the ordinal relations existing amongst trading actions. Therefore, elements such as *time* are excluded in the vector. In addition, *context* and *place* refer to the market and its orderbook which are ignored in the vector, *price* and *volume* are corresponding elements of *goal* and *belief* in the behavioral model respectively, *plan* refers to the probability of trading, *status* refers to trading status, and finally *associate* refers to the number of follow-up trading actions. In this example, we ignore the behavior attributes *plan* and *constraint*, while the *impact* ($f$) of a trading behavior will be further modeled in microstructure behavior pattern analysis in Section 5.3. As a result, the generic abstract behavioral model $\vec{\gamma}$ is instantiated into the following microstructure trading behavior vector.

$$\begin{aligned}\vec{\gamma} &= \{s, o, g, b, a, l, f, u, m\}\\ &= \{account_id, security, price, volume,\\ &trading_action,\ trading_probability,\ impact,\ trading_status,\\ &followup_actions\}\end{aligned} \quad (6)$$

These elements in the microstructure behavior vector are explained as follows.

- $s$ refers to an investor indicated by *account_id*;
- $o$ refers to the *security* traded by an investor;
- $g$ refers to the *price* an investor wants to reach;
- $b$ refers to the order size (*volume*) conducted by a trading behavior instance, $b \in \{b_H, b_M, b_L\}$ represents *large*, *medium* and *small* orders respectively;
- $a$ reflects the *action* of a trading behavior associated with an order, $a \in \{B, S, K, B_i, A_s\}$, where $B_i$ and $A_s$ represent *bid* and *ask* respectively;
- $l$ stands for the probability that an order is traded by the current trading behavior, $l \in \{l_H, l_M, l_L\}$, where $l_H$, $l_M$ and $l_L$ represent *high*, *medium* and *low* probabilities of an order to be traded respectively;
- $f$ reflects the impact of a trading behavior as measured in terms of business interestingness (see Section 5.3);
- $u$ reflects the trading status (say balance of an order) associated with a behavior sequence at the end of a trading period, $u \in \{u_0, u_{-1}, u_1\}$ represents the sequences of trading behavior lead to a *completely traded order*, an *outstanding order* or a *deleted order*;
- $m$ represents the number of associated trading actions following the creation of an order, $m \in \{m_{-1}, m_0, m_1\}$ represents *none*, *one* or *many* trading actions followed.

The example we show here is to mine for exceptional trading behavior that may indicate abnormal trading in the market. We are particularly interested in the following attributes: trade_action, trading_probability, trading_status, followup_actions, because they are the main features related to the microstructure behavior. We then get the following sub-vector of $\vec{\gamma}'$.

$$\vec{\gamma}' = \{b, a, l, u, m\}$$
$$= \{order\text{-}size, trading\text{-}action, trading\text{-}probability, trading\text{-}status, followup\text{-}actions\} \quad (7)$$

Based on this microstructure behavior vector ($\vec{\gamma}'$), market trading behavior can be modeled into vector-based microstructure behavior sequences ($\vec{\Gamma}$'). A vector-based microstructure behavioral sequence ($\vec{\Gamma}$) consists of sequences of behavioral vectors $\vec{\gamma}$ for an investor within a certain trading period. Such behavioral sequences systematically reflect an investor's intention, trading activities, relationships with other associated behavior instances, and behavior procedure (lifecycle) in a market.

Correspondingly, vector-based microstructure behavior sequences ($\vec{\Gamma}$') can be constructed for an investor as follows.

$$\begin{aligned}\vec{\Gamma}' &= \{\vec{\gamma}'_1, \vec{\gamma}'_2, \ldots, \vec{\gamma}'_j, \ldots\} \\ &= \{\vec{\gamma}'_1(b^1, a^1, l^1, u^1, m^1), \\ &\quad \vec{\gamma}'_2(b^2, a^2, l^2, u^2, m^2), \\ &\quad \ldots \\ &\quad \vec{\gamma}'_j(b^j, a^j, l^j, u^j, m^j), \\ &\quad \ldots\}\end{aligned} \quad (8)$$

With the above microstructure behavior vector, we can convert behavior-related orderbook transactions into vector-based microstructure behavior sequences. For this, we also need to consider investor data such as *account_id* that are needed for filling in vector attributes and constructing sequences. As a result of data conversion from transactions to vector-based behavior sequences, a microstructure orderbook is then transformed into microstructure behavioral sequences that are more suitable for explicit behavior pattern and impact analysis.

### *5.3. Mining Microstructure Behavior Patterns*

With the vector-based microstructure behavior sequences, microstructure behavior patterns can be identified. In order to discover behavior patterns, we first define microstructure behavior impact ($f$) (refer to business interestingness $b_i()$) and technical interestingness ($t_i()$) for pattern analysis.

**Definition 1.** (Microstructure Behavior Impact ($f$)) *$f$ in the microstructure behavioral model indicates the impact of microstructure behavior in the market. One way to model the impact is through the calculation of* Abnormal Return *(AR) of all actions in a behavior sequence within a trading period $\Delta t$, which reflects the return volatility of the behavior in the market.*

$$AR = \frac{std(\ln \frac{p_t}{p_{t-1}})}{\sqrt{t}} \quad (9)$$

*where std is the standard division, $p_t$ is the volume-weighted average price on trading a security at time t.*

$$p_t = \frac{\sum g * b}{\sum b} \quad (10)$$

*g and b refer to the price and volume of all orders placed in the relevant time period.*

Two more metrics are defined to measure the technical interestingness of a microstructure behavior pattern. They are *intentional interestingness* ($I_i$) and *exceptional interestingness* ($I_e$).

**Definition 2.** (Intentional Interestingness ($I_i$)) *$I_i$ reflects an investor's intention to deploy a series of trading actions ($\vec{\gamma}$) toward expected impact and goal.*

$$I_i = Supp_t * \frac{|\vec{\gamma}|}{AvgL_t} \tag{11}$$

*where $Supp_t$ is the support of behavior sequence $\vec{\gamma}$ in the behavioral data within a period $\Delta t$, $|\vec{\gamma}|$ is the sequence length, and $AvgL_t$ is the weighted average length of behavior sequences ($\vec{\gamma}$).*

**Definition 3.** (Exceptional Interestingness ($I_e$)) *$I_e$ reflects how exceptional a pattern presents in a target time period than in the benchmark one.*

$$I_e = \frac{\frac{Supp_t}{AvgL_t} * \sum_{j=1}^{m} \omega_j}{\sum_{j=1}^{m} \left(\frac{Supp_j}{AvgL_j} * \omega_j\right)} \tag{12}$$

*where $\omega_j$ is the weight for the benchmark period j, $Supp_j$ is the support of behavior sequence $\vec{\gamma}$ in the behavioral data within a period j which is the benchmark period, and $AvgL_j$ is the weighted average length of behavior sequence $\vec{\gamma}$ in benchmark period j, there are in total m benchmark periods.*

**Definition 4.** (Exceptional Microstructure Behavior Patterns ($P$)) *$P$ is an exceptional microstructure behavior in the market if it satisfies the following conditions:*

$$P(\vec{\gamma}) \in \vec{\Gamma}; \tag{13}$$

$$I_i \geq I_{i0}; \tag{14}$$

$$I_e \geq I_{e0}; \tag{15}$$

*where $I_{i0}$ and $I_{e0}$ are the respective thresholds defined by domain experts.*

The algorithm for identifying such exceptional trading behavior patterns is as follows.

---

METHOD 1: Mining Exceptional Microstructure Behavior in Market Microstructure behavioral data
INPUT: Microstructure behavioral data $\vec{\Gamma}$, thresholds $m$, $I_{i0}$ and $I_{e0}$
OUTPUT: Microstructure behavior patterns $\tilde{P}$

Step 1: Mining general patterns $P$;
FOR $j = 1$ to $m$
Construct the behavior sequences $\vec{\Gamma}_j$ ;
Extract behavior pattern set $P_j$ on $\vec{\Gamma}_j$
ENDFOR
Step 2: Extracting deliverables $\tilde{P}$;
FOR $j = 1$ to $(count(P))$
IF satisfy conditions of $I_i$ and $I_e$ as defined in Definition 4;

Extract the behavior pattern set $\widetilde{P}$;
ENDIF
ENDFOR
Step 3: Converting the patterns $\widetilde{P}$ into business alert rules $\widetilde{R}$.

*5.4. Experiments*

Substantial experiments have been conducted on orderbook data in a large Asian stock exchange. The experimental results shown in this paper come from a dataset consisting of 240 trading days from 2005 to 2006 for a security, which includes 213,898 orders and 228,186 trades. In terms of the microstructure behavior vector model, we convert the orderbook data into vector-based microstructure behavior sequences as defined in Section 5.2. For instance, the following vector-based behavioral sequences are extracted from two consecutive transactions that took place on July 16, 2004.

$$\begin{aligned}\vec{\Gamma}' &= \{\vec{\gamma_1}', \vec{\gamma_2}'\} \\ &= \{(b_L, B, l_L, u_{-1}, m_1), \\ &\quad (b_M, S, l_H, u_1, m_0)\}\end{aligned} \tag{16}$$

We further mine for patterns on the vector-based behavioral sequences. We evaluate all identified patterns based on the following principles: whether a pattern is not only of technical significance but also of business interest. We check all patterns based on:

(1) Technical interestingness: the extent to which they are of high intentional interest ($I_i$) and exceptional interest ($I_e$), which reflect an investor's intention and goal in achieving exceptional performance, and to what extent it can achieve the exceptional performance; and

(2) Business interestingness: the extent to which they lead to high *abnormal return* (*AR*), which indicates exceptionally high return as a result of such trading behavior compared to other normal trading.

Thresholds for the above metrics are defined and refined by involving domain expert knowledge and large experimental findings. Those patterns satisfying both categories of interestingness are selected into the resulting pattern set. The patterns are believed to be not only of a certain level of technical significance, but also of a certain degree of business impact. They are actionable patterns. Table 2 illustrates some exceptional microstructure behavior patterns.

Compared to price-centered exceptional trading pattern analysis, as far as we know, such microstructure behavior patterns disclose much more informative and deeper knowledge about interior trading activities and processes. They also inform the resulting status and investment impact of following these patterns to trade in the market. Such information can assist market surveillance officers to deeply understand market price movement and macroeconomic dynamics with the in-depth driving forces and processes of related investors.

Table 2: Selected Exceptional Microstructure Behavior Patterns

| Date | Exceptional Microstructure Patterns | $I_i$ | $I_e$ | $AR$ (%) |
|---|---|---|---|---|
| 18/01/2005 | $\{(b_S, S, l_M, u_1, m_{-1}),(b_S, S, l_M, u_1, m_{-1}),(b_S, S, l_M, u_1, m_{-1}),(b_S, S, l_M, u_0, m_0)\}$ | 0.026 | 10.8 | 1.85 |
| 01/02/2005 | $\{(b_S, B, l_H, u_1, m_{-1}),(b_S, B, l_H, u_1, m_{-1}),(b_S, B, l_H, u_1, m_{-1})\}$ | 0.030 | 5.2 | 0.93 |
| 24/05/2005 | $\{(b_S, S, l_M, u_0, m_0),(b_S, S, l_M, u_0, m_0)\}$ | 0.054 | 11.2 | 6.38 |
| 13/07/2005 | $\{(b_S, B, l_H, u_{-1}, m_{-1})\}$ | 0.028 | 5.6 | 9.12 |
| 15/09/2005 | $\{(b_S, B, l_H, u_{-1}, m_{-1}),(b_S, B, l_H, u_{-1}, m_{-1})\}$ | 0.035 | 8.8 | 3.49 |
| 05/12/2005 | $\{(b_S, S, l_M, u_0, m_0),(b_S, S, l_M, u_0, m_0)\}$ | 0.035 | 8.6 | 3.41 |

## 6. Case Study 2: Social Security Behavior Analysis

### *6.1. Debt-Related Interaction and Intervention Activities in the Social Security Area*

In the social security network, all entitled customers contact government departments or agencies to obtain government benefits. For various reasons, some benefit recipients intentionally or unintentionally receive overpayments from these organizations. This results in governmental debt. During the debt generation process and the recovery and prevention of the debt lifecycle, there will usually be many actions taken by relevant customers. Once a debt is triggered, intervention actions will be taken by an organization's officers to inspect and recover the debt, or to prevent further debt being incurred. For instance, an officer may call a debtor to arrange a review of his/her debts. As a result, frequent government-customer contacts generate huge quantities of debt-related interaction and intervention activities in the transactions. For example, the Australian social security government agency accumulated a total of 5.4 billion transactions in the financial year 2004-2005 [49]. This data involves 6.5 million people, or one-third of Australians, 2.8 million new claims, 6,600 home visit reviews, 33 million phone calls, and 40 million Internet accesses. Such activities may happen sequentially or in parallel in the business process. The activities are associated with 1.5 million debts (excluding debts of Family Tax Benefit and Child Care Allowance) amounting to AUD$2 billion per annum.

These activity-related transactions consist of information about activity subjects and objects, their actions and follow-ups. It is of high business interest to observe possible indicators hidden in debt-related activities associated with debt occurrences or resolution, as well as the impact of the hidden indicators on debts. For this purpose, we need to extract those debt-related activities from the generic transactions, and re-present them to highlight and analyze the behavior of both the debtor and the corresponding officer in debt recovery and prevention.

### *6.2. Modeling and Constructing Social Security Activity Sequences*

Let us illustrate the idea of activity modeling and construction by scrutinizing activities accumulated in the Australian social security government agency. First, supported by domain analysts, we identify those activity-related transactions, and then extract activities related to debt. Such activities are recoded into activity codes representing interactions and intervention actions which have taken place between customers and officers. As a result, we acquire customer activity data corresponding to their interaction with the Government; Table 3 following illustrates a few such activity codes.

Besides the activity data, we also extract the customer demographic data. Following the vector-based behavioral data model, we build the following vector ($\vec{\gamma}$) to represent debt-related customer demographic data ($\vec{\gamma_1}$) and interaction and intervention activities ($\vec{\gamma_2}$) in the social security area.

Table 3: Activity Codes

| Activity code | Description |
|---|---|
| DOC | Debt-related document |
| REV | Debt-related review of customers |

$$\vec{\gamma} = \{\vec{\gamma_1}, \vec{\gamma_2}\} \tag{17}$$

Unlike the vector used in Case Study 1, here human demographic attributes are also captured and stored into a sub-vector ($\vec{\gamma_1}$), while those related to activities are modeled into another vector ($\vec{\gamma_2}$). The combination of demographic and activity data can present a complete understanding of debt causes and effects compared to demographic-only data.

$\vec{\gamma_1}$ represents the demographic attributes of a customer. It consists of attributes such as *indigenous code* and *rent type* for all customers. The last element in $\vec{\gamma_1}$ is a 'label' indicating whether a customer is likely to get into debt or not. If 'label' = 'DET' indicating a debt, otherwise non-debt with 'NDT'. Such label-associated demographic data is useful for identifying risk demographic factors associated with debt occurrences.

$$\vec{\gamma_1} = \{Person_ID, partner\ Person_ID, indigenous_code, medical_condition, region_office, gender, age, marital_status, birth_country, migration_status, education_level, postcode, language, rent_type, method_of_payment\} \tag{18}$$

The following example illustrates the demographic information of certain customers.

$$\{gender = \text{'}female\text{'}, age = \text{'}65+\text{'}, region_office = \text{'}remote\text{'}, marital\text{-}status = \text{'}separated\text{'}, debt\} \tag{19}$$

$\vec{\gamma_2}$ indicates the activities related to overpaid customers. The vector model used to represent social security activities is as follows. It consists of all activities (activity code) captured within a time window and associated with a particular customer, and the last element is the 'label' indicating the impact of those activities. If 'label' = 'DET' indicating a debt associated with the activities, otherwise non-debt with 'NDT' associated with the activities.

$$\vec{\gamma_2} = \{author, activity_code_1, activity_code_2, \ldots, activity_impact, label\} \tag{20}$$

The following examples show an activity sequence (composed of activity codes such as 'EAN') that happened in January to March 2006. The first is associated with debt (labeled by 'DET'; the debt amount is \$315, and the debt existed for 14 days). The second sequence is associated with non-debt (labeled by 'NDT').

$$\{DOC, EAN, NDB, CCO, STM, EAN, \$315, 14-days, DET\}$$
$$\{REA, STM, AVC, DBT, REA, JSP, CRV, NDT\} \tag{21}$$

As a result of the data construction, we obtain vector-based demographic data ($\overrightarrow{\Gamma_1}$) and vector-based activity sequences ($\overrightarrow{\Gamma_2}$), which consist of the social-security behavioral data ($\overrightarrow{\Gamma} = \{\overrightarrow{\Gamma_1}, \overrightarrow{\Gamma_2}\}$) for further pattern and impact analysis.

### 6.3. *Mining High Impact Behavior Patterns in Social-Security Data*

#### 6.3.1. *Mining Sequential Activity Patterns in Social Security Data*

One exercise is to mine for sequential activity patterns solely on the activity data, and the activities are associated with high business impact (debt in this case) [10]. With the activity sequences constructed, the following types of activity patterns are discovered: (1) Mining Frequent Impact-Oriented Activity Patterns, (2) Mining Sequential Impact-Contrasted Activity Patterns, and (3) Mining Sequential Impact-Reversed Activity Patterns. We briefly introduce them as follows.

1. Mining Frequent Impact-Oriented Activity Patterns. Both *positive* and *negative* activity patterns may be identified that are associated with the behavior impact on business interest. A *positive* pattern has the appearance of activity sequence ($P$), while a *negative* one has the non-appearance of a sequence ($\bar{P}$).
   - Positive Pattern: $P \rightarrow T$ (or $P \rightarrow \bar{T}$), and
   - Negative Pattern: $\bar{P} \rightarrow T$ (or $\bar{P} \rightarrow \bar{T}$)

   Frequent positive and negative impact-oriented activity patterns are helpful for understanding the contribution of an activity sequence appearance or non-appearance to the target impact (namely $T$ and $\bar{T}$).
2. Mining Sequential Impact-Contrasted Activity Patterns. A sequence of activities is associated with two different class labels (say business impacts, $T$ or $\bar{T}$) of different significances ($t_i$) in different situations. For instance, an activity sequence $P$ is associated with label $T$ of very high confidence in dataset $A_T$ but of relatively low confidence with $\bar{T}$ in contrast to $T$ in database $B_{\bar{T}}$.
   - Base Pattern: $P_A \rightarrow T$ indicating the sequence $P$ is associated with impact $T$ in dataset $A_T$;
   - Contrast Pattern: $P_B \rightarrow \bar{T}$ indicating $P$ is associated with impact $\bar{T}$ in dataset $B_{\bar{T}}$.

   where we have $t_{i,A}(P_A \rightarrow T) >> t_{i,B}(P_B \rightarrow \bar{T})$, or $t_{i,A}(P_A \rightarrow T) \quad t_{i,B}(P_B \rightarrow \bar{T})$. This means there is a significant class difference existing between the pattern confidences in different datasets. To identify and measure the class difference of impact-contrasted patterns, the following interestingness metrics are defined to reflect not only class difference, but the likelihood of the occurrences of one impact-oriented pattern against the other types of pattern, or vice versa.

   **Definition 5.** *(Class Difference of Impact-Contrasted Patterns.) The* class differences *of impact-contrasted activity sequence P in two datasets $A_T$ and $B_{\bar{T}}$ are defined as:*

   $$Cd_{T,\bar{T}}(P) = Supp_{A_T}(P \rightarrow T) - Supp_{B_{\bar{T}}}(P \rightarrow \bar{T}) \quad (22)$$

   $$Cd_{\bar{T},T}(P) = Supp_{A_{\bar{T}}}(P \rightarrow \bar{T}) - Supp_{A_T}(P \rightarrow T) \quad (23)$$

   **Definition 6.** *(Class Difference Ratio of Impact-Contrasted Patterns.) The* class difference ratios *of impact-contrasted activity sequence P in $A_T$ and $B_{\bar{T}}$ are defined as:*

$$Cdr_{T,\bar{T}}(P) = \frac{Supp_{A_T}(P \rightarrow T)}{Supp_{B_{\bar{T}}}(P \rightarrow \bar{T})} \tag{24}$$

$$Cdr_{\bar{T},T}(P) = \frac{Supp_{B_{\bar{T}}}(P \rightarrow \bar{T})}{Supp_{A_T}(P \rightarrow T)} \tag{25}$$

Such an impact-contrasted pattern pair is very useful for understanding the different outcomes of doing the same things under varying situations, and guides the behavior impact toward preferred results.

3. Mining Sequential Impact-Reversed Activity Patterns. There are two patterns: one is *underlying*, and the other is *derivative* of the underlying one. They are associated with different class labels $T$ and $\bar{T}$ (in business situations, these two labels may indicate two opposite impacts, say, leading to debt ($DET$) or non-debt ($NDT$) in the social security area).
   - Underlying Pattern: $P \rightarrow T$, which means that activity sequence $P$ leads to target $T$;
   - Derivative Pattern: $PQ \rightarrow \bar{T}$, which means that activity sequence $P$ followed by $Q$ leads to non-target $\bar{T}$.

   Another scenario of impact-reversed activity pattern is the reversal from negative impact-targeted activity pattern $P \rightarrow \bar{T}$ to positive impact $PQ \rightarrow T$ after joining with a trigger activity or activity sequence $Q$.

We further define the following interestingness metrics for impact-reversed activity patterns.

**Definition 7.** *(Conditional Impact Ratio of Impact-Reversed Patterns.) The conditional impact ratio (Cir) is proposed to measure the significance of activity Q-sequence leading to impact reversal from positive to negative or vice versa:*

$$\begin{aligned} Cir(Q\bar{T}|P) &= \frac{Prob(Q\bar{T}|P)}{Prob(Q|P) \times Prob(\bar{T}|P)} \\ &= \frac{Prob(PQ \rightarrow \bar{T})/Prob(P)}{(Prob(PQ)/Prob(P)) \times (Prob(P \rightarrow \bar{T})/Prob(P))} \\ &= \frac{Prob(PQ \rightarrow \bar{T})/Prob(PQ)}{Prob(P \rightarrow \bar{T})/Prob(P)} \end{aligned} \tag{26}$$

*Cir* indicates the impact of $Q$ on the transfer from one type of impact to another. If *Cir* is greater than a given threshold, then it is $Q$ that makes a significant contribution to the change of pattern impact from $P \rightarrow T$ to $PQ \rightarrow \bar{T}$ or vice versa.

**Definition 8.** *(Conditional Piatetsky-Shapiro's Ratio of Impact-Reversed Patterns.) Conditional Piatetsky-Shapiro's Ratio (Cps) measures the difference led by the occurrence of Q (see* Piatetsky-Shapiro's *in [40]), which is defined as follows.*

$$\begin{aligned} Cps(Q\bar{T}|P) &= Prob(Q\bar{T}|P) - Prob(Q|P) \times Prob(\bar{T}|P) \\ &= \frac{Prob(PQ \rightarrow \bar{T})}{Prob(P)} - \frac{Prob(PQ)}{Prob(P)} \times \frac{Prob(P \rightarrow \bar{T})}{Prob(P)} \end{aligned} \tag{27}$$

*Cps* measures the statistical or proportional significance of activity sequence $Q$ leading to the impact reversal.

The impact-reversed pattern pair is very useful for debt prevention action-taking. As long as we know what a magic *Q-sequence* is, business analysts can then either induce or block it to convert the outcome of customer behavior from negative to positive.

In addition, to measure the impact of social-security activity patterns on overpayment occurrences, we build the following risk metrics: $risk_{amt}$ and $risk_{dur}$ to reflect the risk of a pattern leading to debt occurrences in terms of *debt amount* and *debt duration* (taking $P \rightarrow T$ as an example).

$$risk_{amt}(P \rightarrow T) = \frac{\sum_1^{|P\rightarrow T|} d_amt()_i}{\sum_1^{|P|} d_amt()_i} \tag{28}$$

$$risk_{dur}(P \rightarrow T) = \frac{\sum_1^{|P\rightarrow T|} d_dur()_i}{\sum_1^{|P|} d_dur()_i} \tag{29}$$

where pattern $P \rightarrow T$ indicates an activity sequence $P$ is associated with the debt occurrence ($T$), $|P \rightarrow T|$ is the count of pattern $P \rightarrow T$ in the pattern set, $|P|$ is the count of activity sequence $P$ in the dataset, $d_amt()_i$ and $d_dur()_i$ is the debt amount and duration respectively associated with each pattern or activity sequence in the data.

*6.3.2. Mining Demographic-Activity-Combined Patterns in Social Security Data*

The idea of *combined pattern mining* is as follows. Assume that there are $k$ datasets $D_i$ ($i = 1..k$), let $X_i$ be the set of all items in dataset $D_i$ and $\forall i \neq j,\ X_i \cap X_j = \emptyset$. A *combined pattern P* is defined in the form of:

$$P : Y_1 \wedge Y_2 \wedge \ldots \wedge Y_k \rightarrow T, \tag{30}$$

where $Y_i \subseteq X_i$ is an itemset in dataset $D_i$, $T \neq \emptyset$ is a target item or class, and $\exists i, j, i \neq j, Y_i \neq \emptyset, Y_j \neq \emptyset$.

As we have seen from the demographic and activity vectors $(\overrightarrow{\Gamma_1}, \overrightarrow{\Gamma_2})$, both end with a label referring to the same thing ($T$, $T$). Following the idea of combined pattern mining, the demographic-activity-combined patterns are discovered on the two vectors $(\overrightarrow{\Gamma_1}, \overrightarrow{\Gamma_2})$ through pattern merging. The process is as follows.

---

METHOD 2: Demographic-Activity-Combined Patterns Mining
INPUT: target demographic vector $\overrightarrow{\Gamma_1}$ and activity vector $\overrightarrow{\Gamma_2}$
OUTPUT: demographic-activity-combined patterns $\overline{P}$
Step 1: *Demographic pattern mining*: Extracting patterns $D_n$ ($n = 1, \ldots, N$) on demographic vector $\overrightarrow{\Gamma_1}$;
FOR $n = 1$ to $N$
    Develop modeling method $m_{dn}$;
    Employ method $m_{dn}$ on the data $\overrightarrow{\Gamma_1}$
    Extract the demographic pattern set $D_n$;
  ENDFOR
Step 2: *Activity pattern mining*: Extracting activity patterns $A_k$ ($k = 1, \ldots, K$) on activity vector $\overrightarrow{\Gamma_2}$;
  FOR $k = 1$ to $K$
    Develop modeling method $m_{ak}$;

Employ method $m_{ak}$ on the data $\overrightarrow{\Gamma_2}$

Extract the activity pattern set $A_k$;

ENDFOR

Step 3: *Pattern merger*: Extracting demographic-activity-combined patterns $\bar{P}$;

FOR $h = 1$ to $H$

Design the pattern merger functions $\uplus^H P_h$ to merge relevant patterns; Employ the method on $\uplus^H P_h$ the pattern set $D_n$ and $A_k$;

Extract the combined pattern set $P = \uplus^H P_h(D_n, A_k)$;

ENDFOR

Two strategies are developed to merge $\uplus^H P_h$ patterns to demographic-activity-combined pattern pairs and clusters.

1. Combined Pattern Pairs. Assume that $P_1$ and $P_2$ are two combined patterns, their left sides can be split into two parts $D$ and $A$, where $D$ and $A$ are respective itemsets from $\overrightarrow{\Gamma_1}$ and $\overrightarrow{\Gamma_2}$ ($D \neq \emptyset, A \neq \emptyset, D \cap A = \emptyset$). If $P_1$ and $P_2$ share the same $D$ but have different $A$ (or same $A$ but different $D$) and different right sides (namely the impact label), then they consist of a *combined pattern pair* P as

$$\mathcal{P}: \begin{cases} P_1: D \wedge A_1 \rightarrow T_1 \\ P_2: D \wedge A_2 \rightarrow T_2 \end{cases}, \tag{31}$$

where $D \neq \emptyset$, $A_1 \neq \emptyset$, $A_2 \neq \emptyset$, $T_1 \neq \emptyset$, $T_2 \neq \emptyset$, $D \cap A_1 = \emptyset$, $D \cap A_2 = \emptyset$, $A_1 \cap A_2 = \emptyset$ and $T_1 \cap T_2 = \emptyset$.

A combined pattern pair is composed of two contrasting patterns, which suggests that customers with the same demographic characteristics $D$ but different activity sequences $A_1$ and $A_2$, can result in different outcomes, $T_1$ and $T_2$ (say debt or non-debt).

The interestingness of a combined pattern is defined as follows.

**Definition 9 (Interestingness of a Combined Pattern).** *For a combined pattern $P: D \wedge A \rightarrow T$, its interestingness is defined as*

$$I_P(d \wedge a \rightarrow T) = \frac{Cont(D, D \wedge A \rightarrow T)}{Lift(D \rightarrow T)}. \tag{32}$$

*where Cont is the count of relevant patterns.*

$I_P$ measures whether the contribution of $D$ (or $A$) to the occurrence of $T$ increases with $A$ (or $D$) as a precondition. Therefore, "$I_P < 1$" suggests that $D \wedge A \rightarrow T$ is less interesting than $D \rightarrow T$ and $A \rightarrow T$. The value of $I_P$ falls in [0,+ ). When $I_P > 1$, the higher $I_P$ is, the more interesting the rule is. If setting $A = \emptyset$ in Equation 32, then $I_P(D \rightarrow T) = Cont(D, D \rightarrow T)/Lift(D \rightarrow T) = 1$.

Further, the interestingness of a combined pattern pair is defined as follows.

**Definition 10 (Interestingness of Combined Pattern Pair).** *Suppose that* P *is a combined rule pair composed of $P_1$ and $P_2$, the interestingness of the pattern pair* P *is defined as*

$$I_{pair}(\mathrm{P}) = Cont(A, P_1)\ Cont(A, P_2)\ Dist(T_1, T_2), \tag{33}$$

*where $Dist(\cdot)$ denotes the dissimilarity between two descendants and it falls in [0,1]. For binary or nominal attributes, the dissimilarity can be defined as $Dist(T_i, T_j) = 0$ if $T_i = T_j$, otherwise $Dist(T_i, T_j) = 1$.*

$I_{\text{pair}}$ measures the contribution of the two different parts in antecedents to the occurrence of different classes in a group of customers with the same demographics or the same activity patterns. Such knowledge can help to design business campaigns and improve business process. The value of $I_{\text{pair}}$ falls in $[0,+\infty]$. The larger $I_{\text{pair}}$ is, the more interesting and actionable a pair of rules are.

2. Combined Pattern Clusters. Further, based on the combined pattern pair, related combined patterns can be organized into clusters to supplement more information to a pattern pair. A *combined pattern cluster* C is a set of combined patterns based on a combined pattern pair P, where the patterns in C share the same $D$ but have different $A$ on the left side.

$$C : \begin{cases} D \wedge A_1 \rightarrow T_1 \\ D \wedge A_2 \rightarrow T_2 \\ \dots \\ D \wedge A_n \rightarrow T_n \end{cases}, \tag{34}$$

where $D \neq \emptyset$, $\forall i$, $A_i \neq \emptyset$, $T_i \neq \emptyset$, $D \cap A_i = \emptyset$, and $\forall i$ i$= j$, $A_i \cap A_j = \emptyset$.

Based on the interestingness of a combined pattern pair, the interestingness of a combined pattern cluster is defined as follows.

**Definition 11 (Interestingness of Combined Pattern Cluster).** *For a pattern cluster* G *with n combined rules* $P_1, P_2, \dots, P_n$, *its interestingness is*

$$I_{\text{cluster}}(\mathcal{G}) = \max_{i \neq j, P_i, P_j \in \mathcal{G}, T_i \neq T_j} I_{\text{pair}}(P_i, P_j). \tag{35}$$

The above definition of $I_{\text{cluster}}$ indicates that interesting clusters are the rule clusters with interesting rule pairs, and the other rules in the cluster provide additional information. Similar to $I_{\text{pair}}$, the value of $I_{\text{cluster}}$ also falls in $[0,+\infty]$.

In debt recovery and prevention, two types of activities are of particular interest to business. One is the *arrangement* activities that are organized by an officer for a debtor to repay overpayments in terms of payment methods, frequency and amount. Following such an arrangement, *repayment* activities may be undertaken by a debtor to pay the money back to the government. However, the arranged activities may not take place in the repayment period. If this is the case, the debt will sit there. The demographic-activity-combined pattern analysis has the potential to identify relationships among demographics, arrangements and repayments against classes of customers. Such knowledge can inform debt recovery more effectively and quickly.

### *6.4. Experiments*

Substantial experiments have been conducted on social security data to identify high impact behavior patterns related to governmental debts. The data involves customer circumstance data, earnings details, customer-officer interaction activities, and debt details. Activities are categorized into over 200 types of activity codes, such as ‘changing address’, or ‘lodge a document’.

Here we report some of the experimental results in terms of (1) mining sequential activity patterns and (2) mining demographic-activity-combined patterns.

(1). Experimental results on mining sequential activity patterns.

The results shown here come from the data from 1/1/2006 to 31/3/2006. It includes 15,932,832 government-customer interactions with 495,891 customers. Through constructing activity sequences for debt ($T$) and non-debt ($\overline{T}$) as targets by involving domain expert knowledge, 6,063,703 activities, 454,934 activity sequences and 16,540 debt-related sequences are studied. For 5,770,523 income related debt and non-debt activities, 439,953 sequences and 1,559 debt-related sequences are examined.

Table 4: Selected frequent debt-targeted sequential activity patterns ($A$ is the activity dataset)

| Patterns $P \to T$ | $Supp_A(P)$ | $Supp_A(T)$ | $Supp_A(P \to T)$ | Confidence | Lift | $risk_{amt}$ | $risk_{dur}$ |
|---|---|---|---|---|---|---|---|
| $AAI, AVC \to T$ | 0.0018 | 0.0364 | 0.0011 | 0.6222 | 17.1 | 0.037 | 0.008 |
| $AVC, UPD \to T$ | 0.0200 | 0.0364 | 0.0125 | 0.6229 | 17.1 | 0.424 | 0.058 |
| $ANO \to T$ | 0.2613 | 0.0364 | 0.0133 | 0.0511 | 1.4 | 0.362 | 0.370 |
| $UPD \to T$ | 0.1490 | 0.0364 | 0.0162 | 0.1089 | 3.0 | 0.505 | 0.203 |

Table 5: Selected debt-contrasted sequential activity patterns (in separate datasets $A_T$ and $B_{\bar{T}}$)

| Patterns ($P \to T$ in $A_T$, $P \to \bar{T}$ in $B_{\bar{T}}$) | $Cd_{T,\bar{T}}(P)$ | $Cdr_{T,\bar{T}}(P)$ | $Cd_{\bar{T},T}(P)$ | $Cdr_{\bar{T},T}(P)$ | $risk_{amt}$ | $risk_{dur}$ |
|---|---|---|---|---|---|---|
| UPD | 0.309 | 3.24 | -0.309 | 0.31 | 0.505 | 0.203 |
| REA, UPD | 0.309 | 3.60 | -0.309 | 0.28 | 0.487 | 0.184 |
| UPD, REA | 0.227 | 3.12 | -0.227 | 0.32 | 0.376 | 0.163 |
| REA, UPD, DOC | 0.164 | 3.13 | -0.164 | 0.32 | 0.268 | 0.125 |

Table 6.4 illustrates some samples of frequent debt-targeted sequential activity patterns in the activity data. These patterns indicate high risk activities or activity sequences that are not only of high technical significance but also lead to high risk of debt amount and duration.

Table 6.4 illustrates samples of debt-contrasted sequential activity patterns in debt-oriented (dataset $A_T$ with $T$) and non-debt-oriented (dataset $B_{\bar{T}}$ associated with $\bar{T}$) activity datasets respectively. They show that some activities or activity sequences are more likely to lead to debt than non-debt, while others take place in the opposite way.

Table 6.4 shows some results of impact-reversed sequential activity patterns. It is shown that those *Q sequences* play an important role in converting the activity impact from one to the other.

(2). Experimental results on mining demographic-activity-combined patterns

The data used are debts raised in calendar year 2006 and the corresponding customers and arrangements/repayments in the same year. The cleaned sample data contains 355,800 customers with their demographic attributes, arrangements and repayments.

Table 6.4 illustrates some results of demographic-arrangement-repayment-combined patterns associated with certain classes. By combining a pair of similar patterns, we can get much clearer idea of the behavior effects and causes for appearance differences. For instance, it shows that debtors under the same arrangements who also take the same repayment actions may cause different payback effects (classes $A$ and $B$ indicate different payback speed) for different demographic circumstances (even with the same income (both belong to group ‘0’)). For this case, govern-mental debt management officers need to differentiate the customer groups to personalize more effective arrangements for them to repay quickly.

Table 6: Selected impact-reversed sequential activity patterns (in separated datasets $A_T$ and $B_{\bar{T}}$)

| Underlying pattern ($P \to T$) | Derivative pattern ($P, Q \to \bar{T}$) | Cir | Cps |
|---|---|---|---|
| STM | STM,UPD | 2.2 | 0.005 |
| REA | REA | 2.0 | 0.007 |
| DOC | DOC, UPD | 1.7 | 0.006 |
| STM, REA | STM, REA, UPD | 2.0 | 0.006 |
| REA, STM | REA, STM | 2.0 | 0.005 |
| STM, DOC, REA | STM, DOC, REA, DOC | 1.2 | 0.004 |

Table 7: Selected Demographic-Arrangement-Repayment-Combined Pairs

| Patterns | *D* (Demographics) | *A* (Arrangements, | Repayments) | *G* (Class) | *Cont* | *Conf* (%) | $I_P$ | *Lift* | *Lift* of $D \to G$ | *Lift* of $A \to G$ |
|---|---|---|---|---|---|---|---|---|---|---|
| $P_1$ | Income:0 & Remote:Y & Marital:Sep & Gender:F | Withholding | Cash or Post Office & Withholding | B | 20 | 69.0 | 1.47 | 1.95 | 0.91 | 1.46 |
| $P_2$ | Income:0 & Age:65+ | Withholding | Cash or Post Office & Withholding | A | 1123 | 62.3 | 1.38 | 1.35 | 1.24 | 0.79 |

Table 8: Selected Demographic-Arrangement-Repayment-Combined Clusters

| Clusters | Rules | *D* demographics | *A* arrangements | repayments | *G* | *Cont* | *Conf* (%) | $I_P$ | $I_{cluster}$ | *Lift* | *Lift* of $D \to G$ | *Lift* of $A \to G$ |
|---|---|---|---|---|---|---|---|---|---|---|---|---|
| $C_1$ | $p_1$ | age:65+ | withhold | cash or post | A | 1980 | 93.3 | 0.86 | 0.59 | 2.02 | 1.24 | 1.90 |
| | $p_2$ | | irregular | cash or post | A | 462 | 88.7 | 0.87 | | 1.92 | 1.24 | 1.79 |
| | $p_3$ | | withhold & irregular | cash or post | A | 132 | 85.7 | 0.96 | | 1.86 | 1.24 | 1.57 |
| | $p_4$ | | withhold & irregular | withhold | C | 50 | 63.3 | 2.91 | | 3.40 | 0.85 | 1.38 |
| $C_2$ | $p_5$ | benefit:Y | irregular | cash or post | A | 218 | 79.6 | 1.15 | 0.52 | 1.73 | 0.84 | 1.79 |
| | $p_6$ | &age:22-25 | cash | cash or post | C | 483 | 65.6 | 0.78 | | 3.53 | 1.78 | 2.56 |

Table 6.4 illustrates some resulting demographic-arrangement-repayment-combined pattern clusters. Compared to combined pattern pairs, such combined clusters present further detailed indicators for governmental debt management officers to take particular actions on specific target customers in order to convert those slow payers to moderate or even quick payers (as indicated by classes A to C).

## 7. Conclusions

The effective understanding and analysis of behaviors in businesses and complex systems, including virtual organizations play a critical role in disclosing interior driving forces and causes of business problems and appearance. However, in existing management information systems, behavior is implicitly and separately recorded in transactional data. Correspondingly, current so-called behavior analysis is mainly conducted on demographic and service usage data. The resulting outcomes, which focus on the exterior features of business problems, cannot effectively and explicitly scrutinize behavior patterns and impacts on businesses. In this paper, we propose the approach of Behavior Informatics (BI) to study effective methodologies and techniques for explicit and in-depth understanding and analysis of genuine behavioral actions, operations and events associated with many challenging business problems, for instance, the exceptional behavior analysis of terrorists and criminals, and for deep understanding and effective use of interaction and behavior emergence.

Starting from the definition of behavior and a behavioral model, we describe the concept, research issues, process and technical underpinnings of BI. These aspects form a research map of BI, which is worthwhile for further development. To illustrate the concept and its application of BI, we further introduce two real-world case studies: one on market microstructure behavior analysis, and the other on social-security behavior analysis. Substantial experiments in capital markets and social security areas have clearly shown that BI is important, feasible and effec- tive for deeply understanding behavior-oriented problems, and for presenting behavior-oriented solutions that cannot be achieved based on existing approaches. BI has the potential to greatly complement classic analytical approaches, leading to more comprehensive and in-depth business understanding and problem-solving.

Even though relevant knowledge and techniques, such as statistics for behavior analysis, can definitely benefit BI development, there are many BI open issues in aspects such as behavior modeling, behavior simulation, behavior impact analysis, behavior pattern analysis, and behavior presentation. BI is very promising in delivering innovative and effective methodologies and techniques for handling critical business problems in opportunity use and exceptional behavior analysis.

From a scientific field perspective, as any newly-addressed field shows, further studies have to be conducted on the completeness, sufficiency and effectiveness of the proposed framework and potential solutions. Due to the problem's complexity, a practical and effective way is to conduct concrete case studies on real-life scenarios and cases first, and then to summarize the lessons, experiences and knowledge to contribute to a complete and mature community [5] [5].

## 8. Acknowledgment

Thanks are given to Dr Yanchang Zhao, Dr Huaifeng Zhang, Mr Yuming Ou and Ms Shan-shan Wu for their comments and/or experiments on the projects. The work is partially sponsored by Australian Research Discovery Grant (DP1096218, DP0988016 and DP0773412) and Linkage grant (LP0989721 and LP0775041).

## References

[1] Aggarwal, C., Yu, P. Outlier detection for high dimensional data, ACM SIGMOD Record, 30(2), 37-46, 2001.

[2] Au, W., Chan, K., Yao, X. A novel evolutionary data mining algorithm with applications to churn prediction, IEEE Transactions on Evolutionary Computation, 7(6): 532-545, 2003.

[3] Bluhm, C., Overbeck, L., Wagner, C. An introduction to credit risk modeling, Chapman & Hall/CRC, Boca Raton, FL, 2003.

[4] Cao, L., Dai, R., Zhou, M. Metasynthesis: m-space, m-interaction and m-computing for open complex giant systems, IEEE Transactions On Systems, Man, and Cybernetics–Part A, 39(5): 1007-1021, 2009.

[5] Cao, L. Behavior informatics and analytics: let behavior talk, Workshop on Domain Driven Data Mining joint with 2008 International Conference on Data Mining, 87-96, IEEE Computer Society Press, 2008.

[6] Cao, L., Zhang, C. The evolution of KDD: towards domain-driven data mining. International Journal of Pattern Recognition and Artificial Intelligence, 21(4): 677-692, 2007.

[7] Cao, L., Zhang, C., Dai, R. Organization-Oriented Analysis of Open Complex Agent Systems. International Journal of Intelligent Control and Systems, 10(2): 114-122, 2005.

[8] Cao, L., Zhang, C., Liu, J. Ontology-based integration of business intelligence, Web Intelligence and Agent Systems, 4(3): 313-325, 2006.

[9] Cao L., Zhao, Y., Figueiredo, F., Ou, Y., Luo, D. Mining high impact exceptional behavior patterns, Industry Track with 2007 Pacific-Asia Conference on Knowledge Discovery and Data Mining, LNCS4819, 56-63, 2007.

[10] Cao, L., Zhao, Y., Zhang, C. Mining impact-targeted activity patterns in imbalanced data, IEEE Transactions on Knowledge and Data Engineering, 20(8): 1053-1066, 2008.

[11] Cao, L., Zhao, Y., Zhang, C., Zhang, H. Activity mining: from activities to actions, International Journal of Information Technology & Decision Making, 7(2): 259 - 273, 2008.

[5] BI-SIG: http://www.behaviorinformatics.org/

[12] Cao L., Ou, Y. Market microstructure patterns powering trading and surveillance agents. Journal of Universal Computer Sciences, 14(14): 2288-2308, 2008.

[13] Cao, L., Dai, R., Gorodetski, V. Proceedings of Engineering Open Complex Systems - Metasynthesis of Computing Paradigms (EOCS-MCP 2008) Workshop, joint with 2008 IEEE Signature Conference on Computers, Software, and Applications, IEEE Computer Press, 2008.

[14] Cao, L. Integrating agent, service and organizational computing. International Journal of Software Engineering and Knowledge Engineering, 18(5): 573-596, 2008.

[15] Cao L. Activity mining: challenges and prospects. 2006 International Conference on Advanced Data Mining and Applications, LNAI4093, 582-593, 2006.

[16] Carley, K. Artificial Social Agents. http://www.hss.cmu.edu/departments/sds/faculty/carley/publications.htm, 2001.

[17] Chen, H., Chung, W., Xu, J., Wang, G., Qin, Y., Chau, M. Crime data mining: a general framework and some examples, Computer, 37(4): 50-56, 2004.

[18] Curry, C., Grossman, R., Locke, D., Vejcik, S., Bugajski, J. Detecting changes in large data sets of payment card data: a case study, 2007 ACM SIGKDD Conference on Knowledge Discovery and Data Mining, 1018-1022, 2007.

[19] Dasgupta, K., Singh, R., Viswanathan, B., Chakraborty, D., Mukherjea, S., Nanavati, A., Joshi, A. Social ties and their relevance to churn in mobile telecom networks, 2008 International Conference on Extending Database Technology, 668-677, 2008.

[20] Dey, A. Understanding and using context, Personal and Ubiquitous Computing, 5(1):4-7, 2001.

[21] Donoho, S. Early detection of insider trading in option markets, 2004 ACM SIGKDD Conference on Knowledge Discovery and Data Mining, 420-429, 2004.

[22] Eagle, N., Pentland, A., Lazer, D. Inferring social network structure using mobile phone data, Proceedings of the National Academy of Sciences, 106(36): 15274-15278, 2009.

[23] Facca, F., Lanzi, P. Mining interesting knowledge from weblogs: a survey, Data & Knowledge Engineering, 53(3): 225-241, 2005.

[24] Fast, A., Friedland, L., Maier, M., Taylor, B., Jensen, D., Goldberg, H., Komoroske, J. Relational data pre-processing techniques for improved securities fraud detection, 2007 ACM SIGKDD Conference on Knowledge Discovery and Data Mining, 941-949, 2007.

[25] Fawcett, T., Provost, F. Activity monitoring: noticing interesting changes in behavior, 1999 ACM SIGKDD Conference on Knowledge Discovery and Data Mining, 53-62, 1999.

[26] Flesca, S., Greco, S., Tagarelli, A., Zumpano, E. Mining user preferences, page content and usage to personalize website navigation, World Wide Web, 8(3): 317-345, 2005.

[27] Harris, L. Trading and exchanges: market microstructure for practitioners, Oxford University Press, 2003.

[28] Hodge, V., Austin, J. A survey of outlier detection methodologies, Artificial Intelligence Review, 22(2): 85-126, 2004.

[31] Jung, J. Contextualized mobile recommendation service based on interactive social network discovered from mobile users, Expert Systems with Applications, 36(9): 11950-11956, 2009.

[30] Jung, J. Collaborative web browsing based on semantic extraction of user interests with bookmarks, Journal of Universal Computer Science, 11(2): 213-228, 2005.

[31] Jung, J. Social grid platform for collaborative online learning on blogosphere: a case study of eLearning@BlogGrid, Expert Systems with Applications, 36(2): 2177-2186, 2009.

[32] Kobsa, A. Generic user modeling systems, User Modeling and User-Adapted Interaction, 11(1-2): 49-63, 2001.

[33] Kwan, I., Fong, J., Wong, H. An e-customer behavior model with online analytical mining for internet marketing planning, Decision Support Systems, 41(1): 189-204, 2005.

[34] Lane, T., Brodley, C. An empirical study of two approaches to sequence learning for anomaly detection, Machine Learning, 51(1): 73-107, 2003.

[35] Madhavan, A. Market microstructure: a survey, Journal of Financial Markets, 3(3): 205-258, 2000.

[36] Mozer, M., Wolniewicz, R., Grimes, D., Johnson, E., Kaushanky, H. Predicting subscriber dissatisfaction and improving retention in wireless telecommunications industry. IEEE Transactions on Neural Networks, 11(3): 690-696, 2000.

[37] Nanavati, A., Gurumurthy, S., Das, G., Chakraborty, D., Dasgupta, K., Mukherjea, S., Joshi, A. On the structural properties of massive telecom call graphs: findings and implications, 2006 ACM Conference on Information and Knowledge Management, 435-444, 2006.

[38] Neville, j., Simsek, O., Jensen, D., Komoroske, J., Palmer, K., Goldberg, H. Using relational knowledge discovery to prevent securities fraud, 2005 ACM SIGKDD Conference on Knowledge Discovery and Data Mining, 449-458, 2005.

[39] Pompian, M. Behavioral finance and wealth management, Wiley, Hoboken, NJ, 2006.

[40] Sageman, M. Understanding terror networks, University of Pennsylvania Press, 2004.

[41] Srivastava, J., Cooley, R., Deshpande, M., Tan, P. Web usage mining: discovery and applications of usage patterns

from Web data, ACM SIGKDD Explorations Newsletter, 1(2): 12-23, 2000.
[42] Stolfo, S., Hershkop, S., Hu, C., Li, W., Nimeskern, O., Wang, K. Behavior-based modeling and its application to email analysis, ACM Transactions on Internet Technology, 6(2): 187-221, 2006.
[43] Vigna, G., Valeur, F., Kemmerer, R. Designing and implementing a family of intrusion detection systems, ACM SIGSOFT Software Engineering Notes, 28(5): 88-97, 2003.
[44] Wang, F.Y., Carley, K.M., Zeng, D., Mao, W.J. Social computing: from social informatics to social intelligence, IEEE Intelligent Systems, 22(2): 79-83, 2007.
[45] Webb, G., Pazzani, M., Billsus, D. Machine learning for user modeling, User Modeling and User-Adapted Interaction, 11(1-2): 19-29, 2001.
[46] Weiss, G., Hirsh, H. Learning to predict rare events in event sequences, 1998 ACM SIGKDD Conference on Knowledge Discovery and Data Mining, 359-363, 1998.
[47] Yamanishi, K., Takeuchi, J. A unifying framework for detecting outliers and change points from non-stationary time series data, 2002 ACM SIGKDD Conference on Knowledge Discovery and Data Mining, 359-363, 676-681, 2002.
[48] Zhang, Z., Salerno, J., Yu, P. Applying data mining in investigating money laundering crimes, 2003 ACM SIGKDD Conference on Knowledge Discovery and Data Mining, 747-752, 2003.
[49] Centrelink. Centrelink annual report 2004-05.
[50] http://reality.media.mit.edu/